\documentclass[conference]{IEEEtran}
\IEEEoverridecommandlockouts
\usepackage{amsfonts}
\usepackage{textcomp}
\usepackage{xcolor}
\usepackage{fancyhdr}
\usepackage{cite}
\usepackage{graphicx}
\usepackage{psfrag}
\usepackage{subfigure}
\usepackage{url}
\usepackage{stfloats}
\usepackage{amsmath}
\usepackage{array}
\usepackage{fancyhdr}
\usepackage{epsfig}
\usepackage{amssymb}
\usepackage{color}
\usepackage{float}
\usepackage{algorithm}
\usepackage{algpseudocode}
\usepackage{balance}
\usepackage{soul}

\def\BibTeX{{\rm B\kern-.05em{\sc i\kern-.025em b}\kern-.08em
    T\kern-.1667em\lower.7ex\hbox{E}\kern-.125emX}}
\begin{document}

\title{Edge AI Empowered Physical Layer Security for 6G NTN: Potential Threats and Future Opportunities}
\author{\IEEEauthorblockN{Hong-fu Chou, Sourabh Solanki, Vu Nguyen Ha, Lin Chen, \dag Sean Longyu Ma,\\ Hayder Al-Hraishawi, Geoffrey Eappen, Jorge Querol, Juan Carlos  Merlano-Duncan, Symeon Chatzinotas}
\IEEEauthorblockA{\textit{Interdisciplinary Centre for Security, Reliability and Trust (SnT), University of Luxembourg, Luxembourg}\\ 
\textit{\dag School of Computer Science, The University of Auckland, New Zealand}\\ 
%Emails: \{hungpu.chou, sourabh.solanki,vu-nguyen.ha, lin.chen, hayder.al-hraishaw,\\geoffrey.eappen, jorge.querol, juan.duncan, symeon.chatzinotas\}@uni.lu}
Corresponding author: Hong-fu Chou Email: hungpu.chou@uni.lu}
%\author[1]{Hong-fu Chou, Sourabh Solanki, Vu Nguyen Ha}
%\author[2]{Sean Longyu Ma}
%\author[1]{Hayder Al-Hraishawi, Jorge Querol, Symeon Chatzinotas}

%\affil[1]{Interdisciplinary Centre for Security, Reliability and Trust (SnT), University of Luxembourg, Luxembourg}
%\affil[2]{School of Computer Science, The University of Auckland, New Zealand}

}
%Email: $\left\{\rm{hungpu.chou,vu\text{-}nguyen.ha,hayder.al\text{-}hraishawi, ,chitra.shukla,luis.garces,jorge.gonzalez,symeon.chatzinotas}\right\}$@uni.lu} }

\maketitle

\begin{abstract}
Due to the enormous potential for economic profit offered by artificial intelligence (AI) servers, the field of cybersecurity has the potential to emerge as a prominent arena for competition among corporations and governments on a global scale. One of the prospective applications that stands to gain from the utilization of AI technology is the advancement in the field of cybersecurity. To this end, this paper provides an overview of the possible risks that the physical layer may encounter in the context of 6G Non-Terrestrial Networks (NTN). With the objective of showcasing the effectiveness of cutting-edge AI technologies in bolstering physical layer security, this study reviews the most foreseeable design strategies associated with the integration of edge AI in the realm of 6G NTN. The findings of this paper provide some insights and serve as a foundation for future investigations aimed at enhancing the physical layer security of edge servers/devices in the next generation of trustworthy 6G telecommunication networks. 
\end{abstract}
\begin{IEEEkeywords}
6G, NTN, edge AI, edge training, federated learning, deep learning, physical layer security.
\end{IEEEkeywords}

\maketitle

\section{Introduction}
The simultaneous development of both artificial intelligence (AI) and edge computing (EC)  \cite{Deng_2020} is transforming how data is processed, enabling real-time, localized decision-making and enhancing the efficiency and capabilities of various applications and industries. In future communication systems, the integration of AI and EC is expected to be seamless and mutually reinforcing, where AI and EC complement each other by enabling more efficient data processing, reducing latency, enhancing privacy, and extending AI capabilities to the edge of networks. This integration is poised to revolutionize various industries and applications, from IoT and smart cities to autonomous systems and telemedicine. 
Interestingly, by seamlessly integrating AI algorithms with EC, the concept of Edge Intelligence (EI) emerges. 
%The interaction underlying EC and AI is highly dynamic and intricate, including a wide range of ideas and inventions that are intricately interconnected. 
Specifically, EI uses distributed computing at the network's periphery to improve service in the final mile of coverage through the utilization of AI via EC \cite{Edge_2019}. However, expanding the boundaries of AI to encompass the network edge presents significant challenges. These challenges mostly revolve around considerations related to quality, expense, and security. 

The research guidelines in \cite{Edge_2019} have effectively outlined by categorizing EI into numerous study ventures, encompassing both ``AI for edge" and ``AI on edge" approaches, with the aim of enhancing the quality of experience. 
In the realm of EI, security is of paramount importance. One key security aspect involves leveraging Physical Layer Security (PLS) techniques to fortify the EI network against eavesdropping and intrusion attempts. Thus, this research study strives to use and investigate the use cases of EI-empowered PLS in the context of 6G non-terrestrial networks (NTNs).
%in order to continue the discovery of the study in \cite{Mitev2023}. Considering PLS approaches is as essential as using standard encryption \cite{Sanenga2020} owing to latency period specifications, adaptability of widespread IoT deployment, and the threat posed by quantum computing. 
The concept of PLS has gained substantial attention as a supplementary measure alongside existing cryptographic methods at higher network layers. PLS methodologies establish secure communications at the fundamental physical layer, capitalizing on the inherent noise and imperfections in wireless communication channels to intentionally degrade the signal quality received by potential eavesdroppers. This proactive approach effectively impedes eavesdropping attempts and safeguards sensitive information within intercepted signals. 

However, conventional PLS has limitations including the fact that it relies on the eavesdropper's wiretap channel being known in advance and necessitating a high rate of data to guarantee privacy. The modern techniques revealed for example that null-space beamforming techniques can be effectively used in conventional MIMO systems for provisioning of PLS. Therefore, incorporating PLS with EI emerges as a pivotal avenue of study geared toward enhancing the security dimension of 6G networks. First, this challenge is solved by the \textit{bottom-up ``AI for edge"} approach as follows. By establishing a comprehensive and secure handling resources framework utilizing Signal to Interference and Noise Ratio (SINR) to effectively identify secure resources, the aforementioned disadvantage of PLS can be solved by the proposed PLS architecture corporate with EI presented in the work of \cite{Zhao2020} to make it more flexible while designing a key-generating technique in the physical layer that can handle the loss of reciprocation. Following the thread of this approach, it is necessary to make a fundamental shift in the architecture of wireless systems, moving from data-oriented communication to task-oriented communication, in order to achieve the goal of attaining rapid and credible information distillation at the edge of the network. Second, more intriguingly, the \textit{top-down decomposition} ``AI on edge" summarizes the task-oriented communication and comprehensive survey in \cite{mostaani2023taskoriented} and extends its security concerns in \cite{sagduyu2023taskoriented} major jeopardized to the edge device by the presence of evasion and Trojan attacks using adversarial machine learning. The preceding discourse \cite{Liu2023} presents the emergence of secure network design based on EI as a means to offer compelling substantiation for the feasibility of developing an integrated framework that is both information-driven and controlled by comprehension, with the goal of developing and enhancing 6G network security. The authors of the EI network models conducted by \cite{Yu2020} draw inspiration from the field of immunology in order to develop a quick reaction mechanism for situation awareness across diverse nodes. This mechanism is designed to counteract attackers that employ actively evolving assault techniques effectively. Edge AI presents a promising opportunity for enhancing connected knowledge by enabling the efficient allocation of transmission resources at the edge of 6G networks. Edge training and inference \cite{Letaief2022} possesses the ability to retain privacy while also attaining the highest level of security and tolerance for faults.

Motivated by the above, the key contribution of this paper lies in its comprehensive examination of future threats at the physical layer of 6G networks, within the context of EI network models and edge training modes. Additionally, we conduct a thorough review of how EI-assisted LPS offers various research paths aimed at addressing these emerging threats through well-defined design strategies. 

The rest of the paper is structured as follows. In Section II, we discuss the potential threats to the physical layer of 6G networks under the framework of EI network models and edge training modes described in \cite{Letaief2022}. In Section III, we review how EI-assisted PLS examines research avenues that effectively tackle these threats by outlining certain design strategies. Finally, the concluding remarks are presented in Section IV.

\section{Potential Threats to  6G-NTN Physical Layer}
\begin{figure*}[htbp]
\centerline{\includegraphics[width=\textwidth]{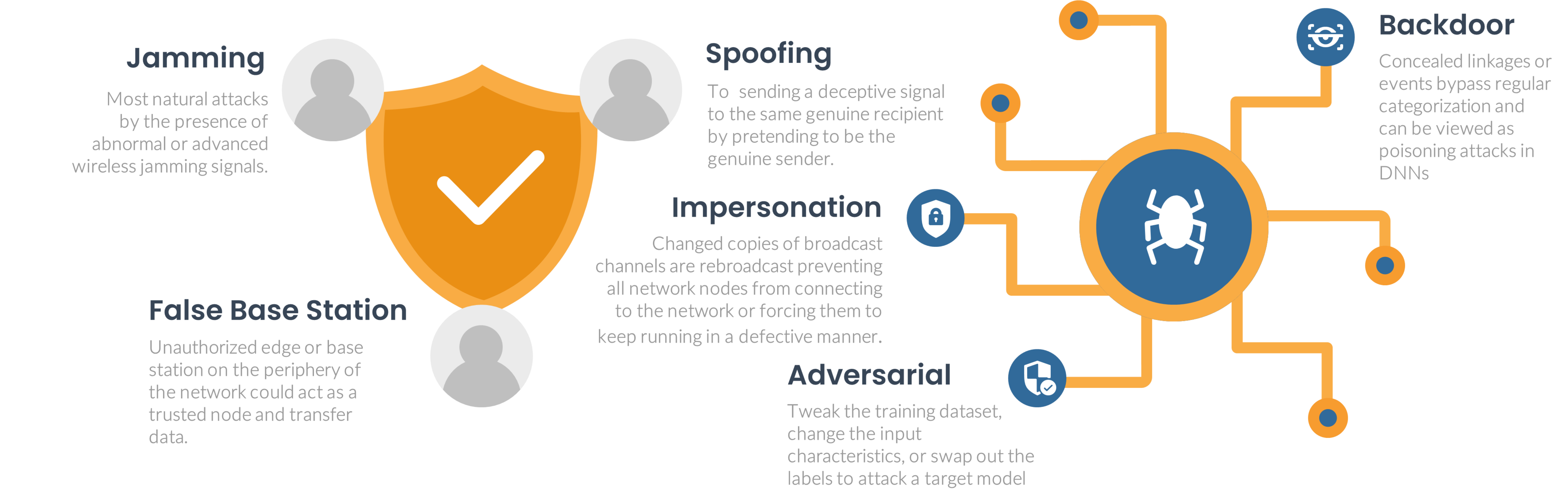}}
\caption{Common Attacks to 6G-NTN physical layer}
\label{edgeai_attacks}
\end{figure*}

The surveys in \cite{li2019physical} and \cite{Tedeschi_2022} prioritized secure operation within the context of 6G NTNs by focusing on security metrics from a PLS viewpoint and conducted a comparative analysis of methods that utilize information-theoretic security techniques, tactics, and schemes. A comprehensive analysis of current study efforts dedicated to incorporating PLS into 6G NTN systems, structured according to three paradigmatic models based on the threats of smart eavesdroppers. An additional security concern pertains to potential denial-of-service (DoS) and distributed denial-of-service (DDoS) attacks\cite{Sharafaldin2019}, wherein adversaries attempt to disrupt satellite or base-station operations by inundating them with an excessive volume of spurious messages. Consequently, satellites targeted in such attacks are compelled to allocate substantial computational resources and time to process these false messages, thereby diminishing the quality of service provided to legitimate users. This susceptibility to DOS attacks is particularly pronounced in the case of LEO satellites due to rather limited computational resources. In the following, we discuss the potential threats to the 6G physical layer, summarized in  Fig.~\ref{edgeai_attacks}.
% We are committed to continuing this research to address the potential threats to the 6G physical layer as follows and summarized in Fig.~\ref{edgeai_attacks}.
\subsection{Jamming Attacks}
Most of the natural attacks in a physical layer can be referred to as instances when the transmission of valid wireless signals is hindered by the presence of abnormal or advanced wireless jamming signals \cite{Pirayesh2021JammingAA}. This interference poses a significant challenge for genuine wireless devices attempting to decode data sequences. Mallory's attack \cite{Mitev2023} as a jammer opts to determine the sensing threshold for a multicarrier system and selectively interfere with subcarriers once signals at power ratings exceeding the specified threshold are detected. In the case of multiple-input multiple-output (MIMO) systems, reliable channel estimation using pilots and the channel state information (CSI) are crucial needs for beamforming in MIMO systems which afterward enables the implementation of precoding techniques. In the event that the CSI is inaccurately assessed or intentionally manipulated by an adversary known as Mallory, the precoder could distribute the power in a scattered manner, leading to possible signal leakage and diminished reliability of the communication channel. Hence, the authors in \cite{Arjoune2020} advise that it is imperative to possess the necessary capabilities for the identification, localization, and elimination of jammers, or alternatively, to deploy strategies for mitigating their effects.

\subsection{Spoofing Attacks}
The spoofer's intention is to trick the actual recipient. Spoofing occurs when a genuine sender of a signal ceases transmitting it to a genuine receiver, at which point the spoofer begins sending a deceptive signal to the same genuine recipient by pretending to be the genuine sender. In \cite{Yılmaz2015}, detection methods and countermeasures to spoofing attacks are discussed. Attempting to jam the spoofer in order to avoid such assaults might cause interference with the other, authentic transceiver in the area due to the scarcity of available frequency bands in the healthcare domain. To circumvent this issue, the transmitter can generate a beam to disrupt the spoofing device without interfering with other transceivers' communications. To damage the transmission without really jamming on the other hand, the spoofer needs just keep sending the transmitter requests again and over. This would cause a breakdown in communication since both transmitters would be sending the same signal. It raises some attention to detect spoofing attacks under mobile users and the
frequency-selective fading channels. 
\subsection{Backdoor and Adversarial Attacks}
Due to its opaque nature, deep neural networks
(DNNs) can be tricked into producing false results via backdoor attacks, in which concealed linkages or events bypass regular categorization and can be viewed as poisoning attacks. To attack a target model, attackers can tweak the training dataset, change the input characteristics, or swap out the labels. This is well-known as a training stage adversarial attack. In \cite{sagduyu2023taskoriented}, the authors use the first scenario to show how adversarial attack on task-oriented communications for 6G NTN networks. The aforementioned phenomenon may be classified as a covert assault, characterized by its elusive nature since it only targets a possibly limited subset of training and test samples. This strategy operates by deceiving the classification process, wherein poisoned test samples in BPSK signal are misidentified as the intended label in QPSK signal. Moreover, whether it's during the gathering of sensing data on the edge device or during sending that data to the gNode, adversarial attacks can be introduced at the DNN's input.
\subsection{False Base Station and Impersonation Attack}
The unauthorized edge or base station on the periphery of the network could act as a trusted node and transfer data to other nodes in the network. Replay by impersonation attacks, in which changed copies of broadcast channels are rebroadcast, can have catastrophic effects on all network nodes, preventing them from connecting to the network or forcing them to keep running in a defective manner. Keys with the properties of mutuality, distinctiveness, and unpredictability of CSI can be generated using the physical layer key generation or RF fingerprint approach. Information is transmitted and received using the same key by both the transmitter and the receiver. However, due to the dynamic nature of the 6G network, changes in the channel are inevitable which makes it hard for the device receiving it to tell if a signal is authentic or not. Therefore, location-based and reverse authentication \cite{Mitev2023} as an open research direction can be achieved through the detection of nearby devices, allowing resistance to these attacks. 
\section{Opportunities and Methodologies of EI-aided PLS}
\begin{figure*}[htbp]
\centerline{\includegraphics[width=\textwidth]{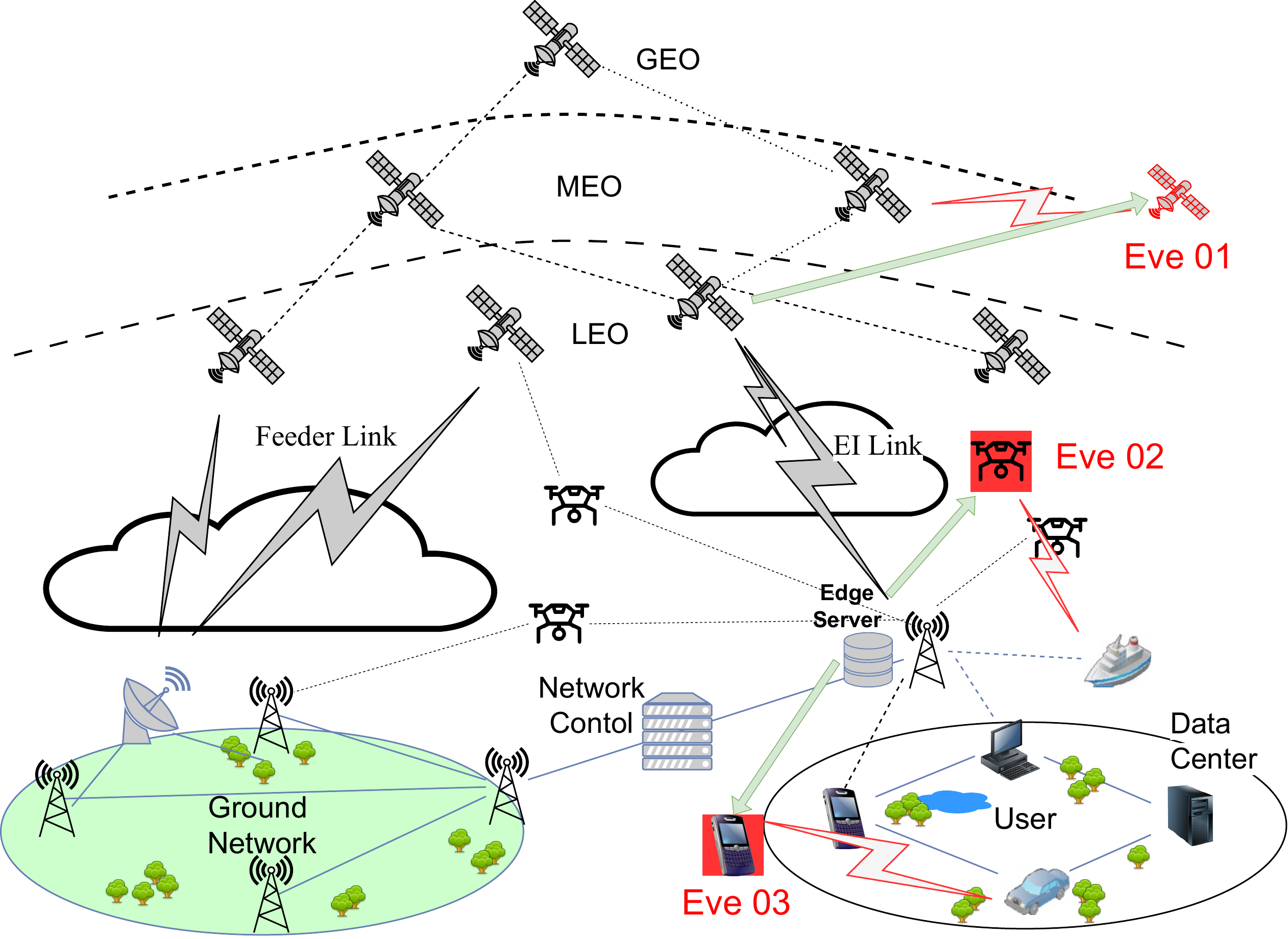}}
\caption{A vision of multi-orbit and multi-layer 6G NTN for PLS}
\label{edgeai_scheme}
\end{figure*}
With the advent of 6G technology, next-generation multiple access schemes and multiple antenna techniques emerge the higher dimensional transmission and drive the application of edge AI. As shown in Fig.\ref{edgeai_scheme}, we present the EI on a multi-orbit multi-layer NTN as a schematic diagram. The ground station and edge server transmits via feeder/EI link to LEO, MEO, and GEO satellite. For the purpose of managing network traffic between user terminals and edge servers/devices, the ground network is coordinated with network control. The diverse wireless network of future 6G NTN \cite{Letaief2022} is enabled by reconfigurable intelligent surface (RIS), cell-free massive MIMO, AirComp, and massive random access to comprehend the significance of edge AI and facilitates the interchange of updates for extremely dimensional frameworks and the adoption of novel network designs. It is intriguing to examine the current advancements and explore the possibilities and approaches of edge AI for physical layer security. In Fig.\ref{edgeai_scheme}, a variety of hackers are illustrated as Eve01 acts as an attacker satellite, Eve02 acts as a malicious drone and Eve03 acts as a hacking user. These potential threats can be mitigated and detected by the following edge AI-empowered PLS design strategies by those green arrows.
\subsection{PLS Design Strategies for Federated Learning}
The constraints imposed by restricted bandwidth and privacy for communication typically render it infeasible to enable all devices at the edge to transmit their data to a central parameter server for the purpose of centralized machine learning \cite{YANG202233}. The introduction of distributed learning algorithms is considered advantageous as it allows devices to collaboratively construct a cohesive learning model through local training and interact with nearby devices. This leads to federated learning (FL) illustrated in Fig.~\ref{edgeai_fl} is seen as a highly promising paradigm to provide design flexibility and still an open research direction of the security aspect to elucidate the benefits of edge AI in relation to PLS concern, for example, the resilience of Byzantine attacks (i.e., a false edge and impersonation attacks to provide malicious updates to the central server) With regard to the decentralized blockchain technology for FL in \cite{MOTHUKURI2021} and edge AI for 6G\cite{Letaief2022}, the authors not only provide a concise overview of the prevailing defensive strategies employed for FL but also a comprehensive study for EI 6G. Therefore, we distill and summarise the PLS strategies for edge AI servers as follows: 
\begin{figure}[htbp]
\centerline{\includegraphics[width=0.5\textwidth]{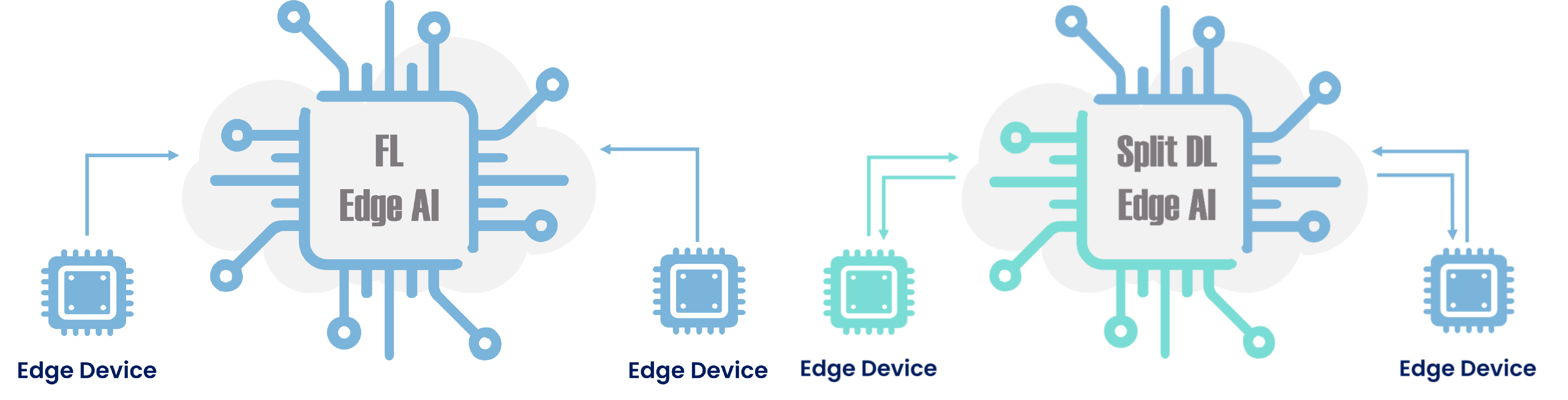}}
\caption{Edge AI applying federal learning and split deep learning}
\label{edgeai_fl}
\end{figure}
\begin{itemize}
\item To mitigate the backdoor and adversarial attacks, the integration of poison-awareness detection\cite{Cao2019,li2020learning} by features to edge AI and distillability of the edge model by transfer learning for local edge privacy\cite{li2019fedmd} are crucial. This design strategy not only assists in verifying an authentic edge server/device but also enhances the level of privacy of the edge AI model throughout the communication process. Furthermore, impact-awareness of edge AI model \cite{Fang2020} is required to delete the training instances at the edge that were tampered with by malicious users. The impact can be measured by either loss function or error rate and merging them both. 
\item As a result of employing the first methodology, the dynamic removal of the edge node occurs regularly in order to address security concerns. This leads to a novel framework of federate multitask learning \cite{Smith2017} which depicts an outlined optimization methodology that takes into account the awareness of systems related to substantial transmission expenses, dawdlers, and tolerance for defects by using lessons learned from the security concern. This intrigues an open research topic requires to investigate. 
\item Trusted execution environment (TEE)\cite{Sabt2015} is essential to consider the design in FL for PLS. Static, dynamic, and semi-dynamic trust are measured and profiled by the metrics of trust function. Leveraging TEE, the authors in \cite{Mo2021,CHEN202069} offer an advanced sense of the aforementioned methodologies to proactively protect privacy by training edge its own data over TEE FL framework and building a TEE to the layer of edge AI in FL framework.
\item For the proactive defense sense, the FoolGold\cite{fung2020mitigating} is proposed as an adapting intelligent methodology for FL by weighting the learning rate of every edge and learning abnormal gradient iterative updates of edge clusters instead of specific symptoms and awareness of attacks in order to maintain FL system performance under a large number of attacks. This approach aims to achieve equilibrium in dealing with adversaries and serves as a source of inspiration for the subsequent research topics.
\item Novel differential privacy for FL\cite{liu2021privacy} measures the edge privacy to determine privacy level in relation to additive noise and disclosure of the collection and analysis of the numerical edge dataset. Beyond a basic static distribution of power assumption, it is important to note the varying influence of additive noise on converging and privacy levels. Combining to optimize resource allocation and edge privacy, further research efforts are necessary to explore this prospective method, taking into consideration various use cases for PLS of 6G NTN. 
\end{itemize}
The aforementioned PLS design methodologies can serve as a reliable FL framework for ensuring the security of 6G NTN based on the architecture of edge server cooperative inference and co-inference with multiple users\cite{Letaief2022}. Moreover, it is worth noting that ultra-reliable and low-latency communication (URLLC) with regard to EI-aided PLS is still an open research challenge of the tradeoff between security and optimal power/latency. Differential privacy for FL\cite{liu2021privacy} and FoolGold\cite{fung2020mitigating} can be the potential candidates of the remedy that can be employed to seize this opportunity. 
\begin{figure*}[]
\centerline{\includegraphics[width=\textwidth]{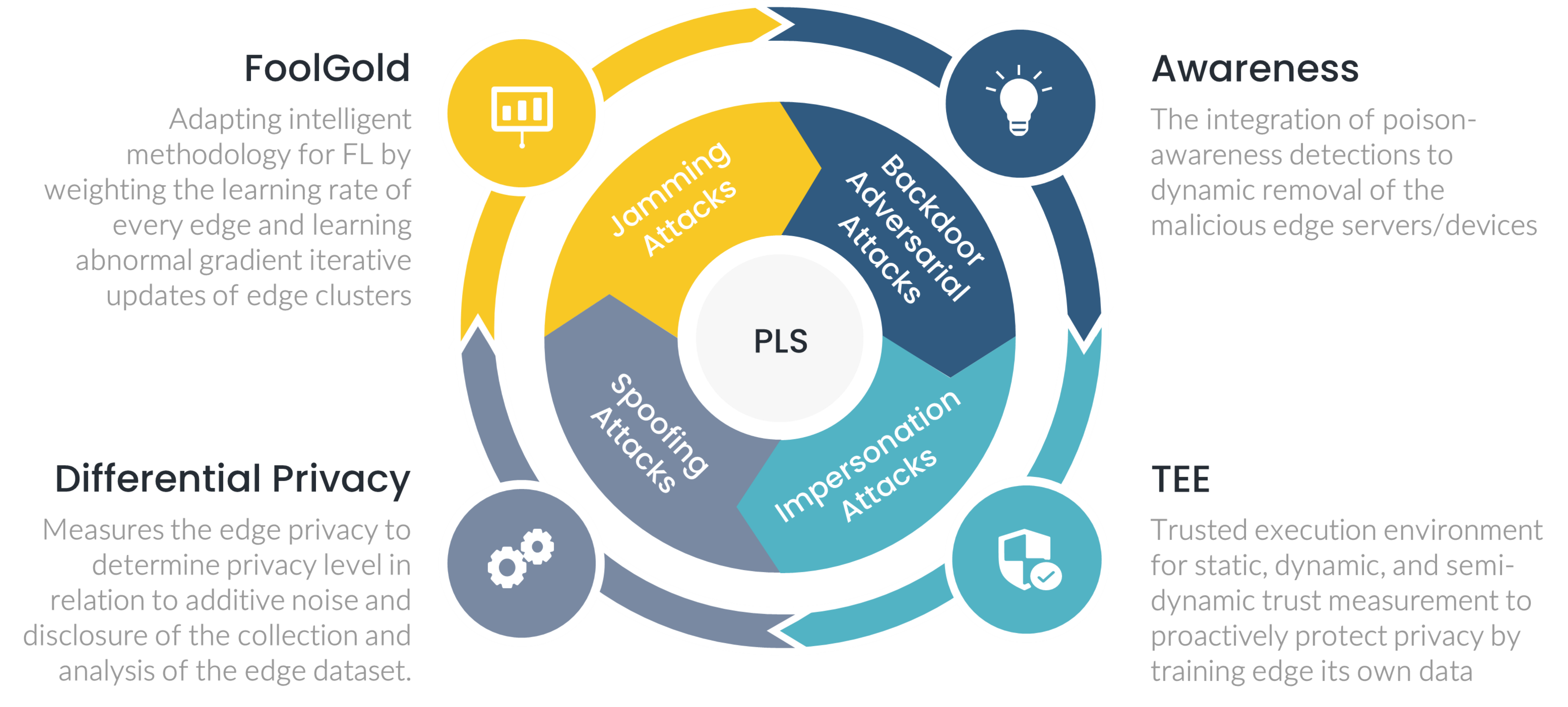}}
\caption{Summary of edge AI empowered PLS}
\label{edgeai_sum}
\end{figure*}
\subsection{Split Deep Learning}
Following the design strategy of FL in the PLS sense, the secure model aggregation concluded in \cite{Letaief2022} can refer to the PLS mentioned above design methodologies. In Fig.~\ref{edgeai_fl}, the authors address that the model split learning of the secure model aggregation is hence especially appropriate for deep learning (DL) with a high number of model parameters. In contrast to the data partition-based training technique, FL necessitates the edge device uploading of an entire replicated global model at each edge device involved. Despite the split DL claims that the objective is to maintain privacy while ensuring computational flexibility, this is achieved by restricting the exchange of primary information and enabling only edge devices to do basic computing for the smaller portions of the edge cluster. The challenge of split DL is still the goal of URLLC with regard to transmission latency within the edge node with limited computing particularly in the instance of a massive user number of 6G NTN edge cluster. However, a study case\cite{abuadbba2020can} reveals the potential method of information leakage due to convertible transmission parameters while training. Therefore, binarized split DL\cite{pham2023binarizing} not only reduces the computation effort for edge device but also provides a methodology to prevent information leakage during the training process by integrating the strategy of differential privacy with the interference compensation by offset Laplace noise. Furthermore, a hybrid split and federal framework is presented in \cite{turina2020combining} to analyze the merits based on the case experiment of privacy-awareness and privacy-oblivious to surpass the split framework. The integration of split deep learning into the task-oriented joint source-channel coding (JSCC) framework for 6G NTN holds significant potential. However, further research is needed to explore the extent of privacy that can be attained in this scenario. We list but are not limited to several points to investigate further:

\begin{enumerate}
\item The transition of secrecy rate between edge servers and devices with JSCC can be examined in the context of differential privacy under an eavesdropping or aforementioned attack scenario. 
\item Split DL latency requirement and resilience measurement with fix-point analysis of task-oriented JSCC framework especially applying binarize scheme in \cite{pham2023binarizing}. 
\item The exploration of various split layers, which are generated in conjunction with source coding, can be further examined by introducing other design parameters and privacy level impacts on the reliability of channel coding.
\item TEE strategy applies to split DL task-oriented JSCC framework to secure partial learning process. The privacy level measurement can be discussed to reveal potential merits.
\item Based on the strategies inspired by FoolGold\cite{fung2020mitigating}, the task-oriented feature learning\cite{Shi2023} enables the potential research direction of malicious feature learning not only to optimize the resource allocation but also monitor the malicious activities by sensing the feature vectors in FL framework with edge split inference.
\end{enumerate}
\subsection{Deep Learning}
The previous iteration of edge AI can be directed towards the mobile edge computing (MEC) network, in light of the rapid advancement of 5G/6G technology. The evolution is to achieve quicker response times and less power usage on the edge device, as compared to traditional methods. This is accomplished by strategically deploying computational access points (CAPs) as edge servers in close proximity to the data sources. The nearby edge devices have the capability to delegate a portion of their computing activities to the nearby CAPs through judicious offloading selections. The researchers in \cite{CHEN2022180} investigate a secure MEC system for PLS in which a network is susceptible to the presence of a number of eavesdroppers. These eavesdroppers pose a potential danger to the process of job offloading. This study proposes an innovative approach that combines deep reinforcement learning with convex optimization to effectively determine a suitable remedy for the offloading proportion, while simultaneously optimizing energy consumption and processing capabilities. Furthermore, the RIS-aided MEC network for 6G NTN is investigated in \cite{zhang2022deep}.  This approach conceals specific contents of the physical layer and offers a streamlined and efficient means of guaranteeing security at the physical layer. In addition to facilitating safe transmission, this strategy guarantees optimal system performance by utilizing deep reinforcement learning techniques identical to \cite{zhang2022deep}. The physical layer authentication can be beneficial from edge AI presented in \cite{li2019physical} to mitigate spoofing attacks on MEC systems. This methodology leverages DNNs for the purpose of facilitating multi-user authentication in order to discern valid edge nodes, hackers, and harmful nodes without reliance on a predetermined threshold for testing. Furthermore, multi-user authentication is accomplished by the selection of cross entropy as the loss function for neural networks and the provision of the vectorization cost function. The research conducted by \cite{li2019physical} demonstrates that the thrilled experimental findings can be extrapolated to future research endeavors in the 6G NTN environment.
\subsection{Experiment Demonstration}
\begin{figure*}[!t]
\centerline{\includegraphics[width=0.9\textwidth]{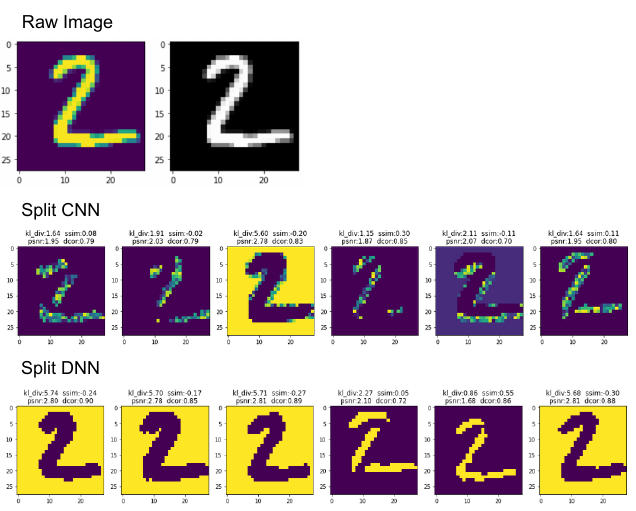}}
\caption{The stimulating imagery of privacy measure using split deep learning in \cite{pham2023binarizing}}
\label{edgeai_split}
\end{figure*}

\begin{figure}[htbp]
\centerline{\includegraphics[width=0.5\textwidth]{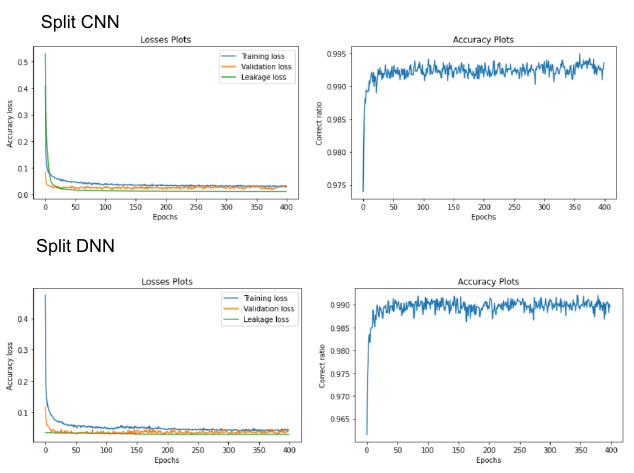}}
\caption{The accuracy loss and correct ratio using split deep learning in \cite{pham2023binarizing}}
\label{edgeai_split2}
\end{figure}

We apply online source code in \cite{pham2023binarizing} to exhibit the potential risks in PLS by demonstrating the privacy leaking impact of the original dataset utilizing split deep learning. In Fig.\ref{edgeai_split}, the convolutional neural networks (CNN) and DNN are divided into edge server and device separately. The influence of Eves or hackers on leakage can be assessed using Kullback-Leibler divergence (KL-D) and shows graphically presented from stimulation at a CNN and DNN split layer with one localized convolution in the edge device. The learning curve under information leakage of KL-D is based on training, validation and leakage loss and is presented in Fig.\ref{edgeai_split2} for split CNN and DNN respectively.

\section{Conclusion}
This study offers a comprehensive overview, depicted in Fig.~\ref{edgeai_sum}, of potential security risks that possess the capability to inflict damage upon the physical layer of 6G NTN.
The figure illustrates how edge AI-empowered PLS can facilitate the implementation of protective measures against prospective risks. The twin rings encircle the main objective of PLS entailing the deployment of strategic approaches that can be dynamically rotated to defend against physical layer attacks originating from malicious hackers. Additionally, we revisit several design methodologies for edge AI that can effectively prevent these threats. These strategies involve the utilization of machine learning techniques as well as the implementation of federated learning approaches for edge nodes of the 6G NTN. The perspectives presented in this paper offer valuable knowledge and establish a basis for inspiring research endeavors focused on improving the physical layer security of edge computing in forthcoming 6G NTN.  
%\section*{Acknowledgment}
%This work has been supported by the Luxembourg National Research Fund (FNR), through the CORE Project TRANTOR.
\bibliographystyle{IEEEtran}
\balance
\bibliography{aiecpls6g}

\begin{thebibliography}{10}
\providecommand{\url}[1]{#1}
\csname url@samestyle\endcsname
\providecommand{\newblock}{\relax}
\providecommand{\bibinfo}[2]{#2}
\providecommand{\BIBentrySTDinterwordspacing}{\spaceskip=0pt\relax}
\providecommand{\BIBentryALTinterwordstretchfactor}{4}
\providecommand{\BIBentryALTinterwordspacing}{\spaceskip=\fontdimen2\font plus
\BIBentryALTinterwordstretchfactor\fontdimen3\font minus
  \fontdimen4\font\relax}
\providecommand{\BIBforeignlanguage}[2]{{%
\expandafter\ifx\csname l@#1\endcsname\relax
\typeout{** WARNING: IEEEtran.bst: No hyphenation pattern has been}%
\typeout{** loaded for the language `#1'. Using the pattern for}%
\typeout{** the default language instead.}%
\else
\language=\csname l@#1\endcsname
\fi
#2}}
\providecommand{\BIBdecl}{\relax}
\BIBdecl

\bibitem{Deng_2020}
S.~Deng, H.~Zhao, W.~Fang, J.~Yin, S.~Dustdar, and A.~Y. Zomaya, ``Edge
  intelligence: The confluence of edge computing and artificial intelligence,''
  \emph{IEEE Internet of Things Journal}, vol.~7, no.~8, pp. 7457--7469, 2020.

\bibitem{Edge_2019}
Z.~Zhou, X.~Chen, E.~Li, L.~Zeng, K.~Luo, and J.~Zhang, ``Edge intelligence:
  Paving the last mile of artificial intelligence with edge computing,''
  \emph{Proceedings of the IEEE}, vol. 107, no.~8, pp. 1738--1762, 2019.

\bibitem{Zhao2020}
L.~Zhao, X.~Zhang, J.~Chen, and L.~Zhou, ``Physical layer security in the age
  of artificial intelligence and edge computing,'' \emph{IEEE Wireless
  Communications}, vol.~27, no.~5, pp. 174--180, 2020.

\bibitem{mostaani2023taskoriented}
A.~Mostaani, T.~X. Vu, S.~K. Sharma, V.-D. Nguyen, Q.~Liao, and S.~Chatzinotas,
  ``Task-oriented communication design in cyber-physical systems: A survey on
  theory and applications,'' \emph{IEEE Access}, vol.~10, pp.
  133\,842--133\,868, 2022.

\bibitem{sagduyu2023taskoriented}
Y.~E. Sagduyu, S.~Ulukus, and A.~Yener, ``Task-oriented communications for
  nextg: End-to-end deep learning and ai security aspects,'' \emph{IEEE
  Wireless Communications}, vol.~30, no.~3, pp. 52--60, 2023.

\bibitem{Liu2023}
\BIBentryALTinterwordspacing
J.~Liu, L.~Bai, C.~Jiang, and W.~Zhang, \emph{Future Trend of Network
  Security}.\hskip 1em plus 0.5em minus 0.4em\relax Singapore: Springer Nature
  Singapore, 2023, pp. 409--425. [Online]. Available:
  \url{https://doi.org/10.1007/978-981-99-1125-7_6}
\BIBentrySTDinterwordspacing

\bibitem{Yu2020}
Q.~Yu, J.~Ren, J.~Zhang, S.~Liu, Y.~Fu, Y.~Li, L.~Ma, J.~Jing, and W.~Zhang,
  ``An immunology-inspired network security architecture,'' \emph{IEEE Wireless
  Communications}, vol.~27, no.~5, pp. 168--173, 2020.

\bibitem{Letaief2022}
K.~B. Letaief, Y.~Shi, J.~Lu, and J.~Lu, ``Edge artificial intelligence for 6g:
  Vision, enabling technologies, and applications,'' \emph{IEEE Journal on
  Selected Areas in Communications}, vol.~40, no.~1, pp. 5--36, 2022.

\bibitem{li2019physical}
B.~Li, Z.~Fei, C.~Zhou, and Y.~Zhang, ``Physical-layer security in space
  information networks: A survey,'' \emph{IEEE Internet of things journal},
  vol.~7, no.~1, pp. 33--52, 2019.

\bibitem{Tedeschi_2022}
\BIBentryALTinterwordspacing
P.~Tedeschi, S.~Sciancalepore, and R.~D. Pietro, ``Satellite-based
  communications security: A survey of threats, solutions, and research
  challenges,'' \emph{Computer Networks}, vol. 216, p. 109246, oct 2022.
  [Online]. Available: \url{https://doi.org/10.1016%2Fj.comnet.2022.109246}
\BIBentrySTDinterwordspacing

\bibitem{Sharafaldin2019}
I.~Sharafaldin, A.~H. Lashkari, S.~Hakak, and A.~A. Ghorbani, ``Developing
  realistic distributed denial of service (ddos) attack dataset and taxonomy,''
  in \emph{2019 International Carnahan Conference on Security Technology
  (ICCST)}, 2019, pp. 1--8.

\bibitem{Pirayesh2021JammingAA}
\BIBentryALTinterwordspacing
H.~Pirayesh and H.~Zeng, ``Jamming attacks and anti-jamming strategies in
  wireless networks: A comprehensive survey,'' \emph{IEEE Communications
  Surveys \& Tutorials}, vol.~24, pp. 767--809, 2021. [Online]. Available:
  \url{https://api.semanticscholar.org/CorpusID:230438804}
\BIBentrySTDinterwordspacing

\bibitem{Mitev2023}
M.~Mitev, A.~Chorti, H.~V. Poor, and G.~P. Fettweis, ``What physical layer
  security can do for 6g security,'' \emph{IEEE Open Journal of Vehicular
  Technology}, vol.~4, pp. 375--388, 2023.

\bibitem{Arjoune2020}
Y.~Arjoune and S.~Faruque, ``Smart jamming attacks in 5g new radio: A review,''
  in \emph{2020 10th Annual Computing and Communication Workshop and Conference
  (CCWC)}, 2020, pp. 1010--1015.

\bibitem{Yılmaz2015}
M.~H. Yılmaz and H.~Arslan, ``A survey: Spoofing attacks in physical layer
  security,'' in \emph{2015 IEEE 40th Local Computer Networks Conference
  Workshops (LCN Workshops)}, 2015, pp. 812--817.

\bibitem{YANG202233}
\BIBentryALTinterwordspacing
Z.~Yang, M.~Chen, K.-K. Wong, H.~V. Poor, and S.~Cui, ``Federated learning for
  6g: Applications, challenges, and opportunities,'' \emph{Engineering},
  vol.~8, pp. 33--41, 2022. [Online]. Available:
  \url{https://www.sciencedirect.com/science/article/pii/S2095809921005245}
\BIBentrySTDinterwordspacing

\bibitem{MOTHUKURI2021}
\BIBentryALTinterwordspacing
V.~Mothukuri, R.~M. Parizi, S.~Pouriyeh, Y.~Huang, A.~Dehghantanha, and
  G.~Srivastava, ``A survey on security and privacy of federated learning,''
  \emph{Future Generation Computer Systems}, vol. 115, pp. 619--640, 2021.
  [Online]. Available:
  \url{https://www.sciencedirect.com/science/article/pii/S0167739X20329848}
\BIBentrySTDinterwordspacing

\bibitem{Cao2019}
D.~Cao, S.~Chang, Z.~Lin, G.~Liu, and D.~Sun, ``Understanding distributed
  poisoning attack in federated learning,'' in \emph{2019 IEEE 25th
  International Conference on Parallel and Distributed Systems (ICPADS)}, 2019,
  pp. 233--239.

\bibitem{li2020learning}
S.~Li, Y.~Cheng, W.~Wang, Y.~Liu, and T.~Chen, ``Learning to detect malicious
  clients for robust federated learning,'' \emph{arXiv preprint
  arXiv:2002.00211}, 2020.

\bibitem{li2019fedmd}
D.~Li and J.~Wang, ``Fedmd: Heterogenous federated learning via model
  distillation,'' \emph{arXiv preprint arXiv:1910.03581}, 2019.

\bibitem{Fang2020}
M.~Fang, X.~Cao, J.~Jia, and N.~Z. Gong, ``Local model poisoning attacks to
  byzantine-robust federated learning,'' in \emph{Proceedings of the 29th
  USENIX Conference on Security Symposium}, ser. SEC'20.\hskip 1em plus 0.5em
  minus 0.4em\relax USA: USENIX Association, 2020.

\bibitem{Smith2017}
V.~Smith, C.-K. Chiang, M.~Sanjabi, and A.~Talwalkar, ``Federated multi-task
  learning,'' in \emph{Proceedings of the 31st International Conference on
  Neural Information Processing Systems}, ser. NIPS'17.\hskip 1em plus 0.5em
  minus 0.4em\relax Red Hook, NY, USA: Curran Associates Inc., 2017, p.
  4427–4437.

\bibitem{Sabt2015}
M.~Sabt, M.~Achemlal, and A.~Bouabdallah, ``Trusted execution environment: What
  it is, and what it is not,'' in \emph{2015 IEEE Trustcom/BigDataSE/ISPA},
  vol.~1, 2015, pp. 57--64.

\bibitem{Mo2021}
\BIBentryALTinterwordspacing
F.~Mo, H.~Haddadi, K.~Katevas, E.~Marin, D.~Perino, and N.~Kourtellis, ``Ppfl:
  Privacy-preserving federated learning with trusted execution environments,''
  in \emph{Proceedings of the 19th Annual International Conference on Mobile
  Systems, Applications, and Services}, ser. MobiSys '21.\hskip 1em plus 0.5em
  minus 0.4em\relax New York, NY, USA: Association for Computing Machinery,
  2021, p. 94–108. [Online]. Available:
  \url{https://doi.org/10.1145/3458864.3466628}
\BIBentrySTDinterwordspacing

\bibitem{CHEN202069}
\BIBentryALTinterwordspacing
Y.~Chen, F.~Luo, T.~Li, T.~Xiang, Z.~Liu, and J.~Li, ``A training-integrity
  privacy-preserving federated learning scheme with trusted execution
  environment,'' \emph{Information Sciences}, vol. 522, pp. 69--79, 2020.
  [Online]. Available:
  \url{https://www.sciencedirect.com/science/article/pii/S0020025520301201}
\BIBentrySTDinterwordspacing

\bibitem{fung2020mitigating}
C.~Fung, C.~J.~M. Yoon, and I.~Beschastnikh, ``Mitigating sybils in federated
  learning poisoning,'' \emph{arXiv preprint arXiv:1808.04866}, 2020.

\bibitem{liu2021privacy}
D.~Liu and O.~Simeone, ``Privacy for free: Wireless federated learning via
  uncoded transmission with adaptive power control,'' \emph{arXiv preprint
  arXiv:2006.05459}, 2021.

\bibitem{abuadbba2020can}
S.~Abuadbba, K.~Kim, M.~Kim, C.~Thapa, S.~A. Camtepe, Y.~Gao, H.~Kim, and
  S.~Nepal, ``Can we use split learning on 1d cnn models for privacy preserving
  training?'' in \emph{Proceedings of the 15th ACM Asia Conference on Computer
  and Communications Security}, 2020, pp. 305--318.

\bibitem{pham2023binarizing}
N.~D. Pham, A.~Abuadbba, Y.~Gao, T.~K. Phan, and N.~Chilamkurti, ``Binarizing
  split learning for data privacy enhancement and computation reduction,''
  \emph{IEEE Transactions on Information Forensics and Security}, vol.~18, pp.
  3088--3100, 2023.

\bibitem{turina2020combining}
V.~Turina, Z.~Zhang, F.~Esposito, and I.~Matta, ``Combining split and federated
  architectures for efficiency and privacy in deep learning,'' in
  \emph{Proceedings of the 16th International Conference on emerging Networking
  EXperiments and Technologies}, 2020, pp. 562--563.

\bibitem{Shi2023}
Y.~Shi, Y.~Zhou, D.~Wen, Y.~Wu, C.~Jiang, and K.~B. Letaief, ``Task-oriented
  communications for 6g: Vision, principles, and technologies,'' \emph{IEEE
  Wireless Communications}, vol.~30, no.~3, pp. 78--85, 2023.

\bibitem{CHEN2022180}
\BIBentryALTinterwordspacing
L.~Chen, S.~Tang, V.~Balasubramanian, J.~Xia, F.~Zhou, and L.~Fan,
  ``Physical-layer security based mobile edge computing for emerging cyber
  physical systems,'' \emph{Computer Communications}, vol. 194, pp. 180--188,
  2022. [Online]. Available:
  \url{https://www.sciencedirect.com/science/article/pii/S0140366422002857}
\BIBentrySTDinterwordspacing

\bibitem{zhang2022deep}
L.~Zhang, S.~Lai, J.~Xia, C.~Gao, D.~Fan, and J.~Ou, ``Deep reinforcement
  learning based irs-assisted mobile edge computing under physical-layer
  security,'' \emph{Physical Communication}, vol.~55, p. 101896, 2022.

\end{thebibliography}
\end{document}